\documentclass[12pt,preprint]{aastex}
\shorttitle{Kea}
\shortauthors{Endl \& Cochran}

\begin{document}

\title{Kea: a new tool to obtain stellar parameters from low to moderate 
signal/noise and high-resolution Echelle spectra.
\footnote{Named after {\it Nestor notabilis} an alpine parrot native to New Zealand}
}

\author{Michael Endl \& William D. Cochran}
\affil{McDonald Observatory and Department of Astronomy, The University of Texas at Austin,  
    Austin, TX 78712, USA}

\begin{abstract}
In this paper we describe {\it Kea} a new spectroscopic fitting method to derive stellar parameters from moderate to low signal/noise, high-resolution
spectra. We developed this new tool to analyze the massive data set of the {\it Kepler} mission reconnaissance spectra
that we have obtained at McDonald Observatory. 
We use {\it Kea} to determine effective temperatures ($T_{\rm eff}$), metallicity ([Fe/H]), surface gravity (log~$g$) and projected
rotational velocity ($v \sin i$). {\it Kea} compares the observations to a large library of synthetic spectra that covers a wide range of different 
$T_{\rm eff}$, [Fe/H] and log $g$ values. 
We calibrated {\it Kea} on observations of well-characterized standard stars (the {\it Kepler} field ``platinum'' sample) which
range in $T_{\rm eff}$ from 5000 to 6500~K, in [Fe/H] from -0.5 to +0.4 dex and in log~$g$ from 3.2 to 4.6 dex.
We then compared the {\it Kea} results from reconnaissance spectra of 45 KOIs (Kepler Object of Interest) 
to stellar parameters derived from higher signal/noise spectra obtained with Keck/HIRES. 
We find typical uncertainties of 100~K in $T_{\rm eff}$, 0.12~dex in [Fe/H] and 0.18~dex in log~$g$. 
\end{abstract}


\section{Introduction}

An important step in the Follow-up Observing Program of NASA's {\it Kepler} mission (Borucki et al.~2010) is the acquisition of reconnaissance 
spectra of Kepler Objects of Interest (KOI), i.e. stars hosting transiting planet candidates. These spectra allow a more detailed 
characterization of the potential planet-hosting star. For most KOIs only photometrically derived properties from the Kepler Input Catalog (KIC) are known 
prior to reconnaissance spectroscopy. KIC uncertainities for dwarf stars between 4500 and 6500~K are $\approx 200$~K and 0.4~dex in log~$g$
and somewhat larger for more evolved stars (Brown et al.~2011). 

Stellar radii are often estimated by comparing a star's effective temperature and surface gravity to evolutionary tracks. However, the KIC does not offer tight 
constraints on log~$g$ (0.4~dex at best). Spectroscopic surface gravity values are often more precise and can yield tighter constraints on stellar radii.
This in turn impacts the planetary radius derived from the modeling of the transit feature in the {\it Kepler} light curve.

A large number of different tools have been developed in order to determine
fundamental stellar parameters from observed spectra.  The success of these
tools depends strongly on the nature of the spectra.  Some techniques work
best on high signal/noise (S/N) data and break down for noisy data.  Others work
well on high resolution spectra, but fail when line-blending becomes
significant in low-resolution data.  Thus, one must carefully match the
technique to the type of spectra that will be obtained.

One class of techniques starts with a model stellar atmosphere and computes
the emergent spectrum, given the atomic (or molecular) parameters
(wavelength, excitation potential, $\log gf$, damping constants, etc.)
of the spectral features.
The current ``gold-standard'' in model atmosphere analysis of stellar
spectra are self-consistent 3-D radiation-hydrodynamics simulations
(e.g. Stein \& Nordlund~1998, Asplund et al.~2000, Magic et al.~2013).
These models succeed quite well in
reproducing the details of stellar line profile shapes.
However, this level of detail really demands spectra
of comparable quality, with very high spectral resolution and S/N.
A widely used ``workhorse'' alternative is the MOOG stellar atmospheric
analysis code (Sneden~1973). MOOG performs a variety of spectral
line analysis and spectrum synthesis tasks under the assumption of local
thermodynamic equilibrium (LTE) in a 1-D stellar atmospheric model.
With MOOG, one can either measure equivalent widths of individual stellar
atmospheric lines of interest and compare those with model widths,
or one can synthesize regions of stellar spectra for comparison with
the observed spectrum.
A third alternative is ``Spectroscopy Made Easy'', or SME (Valenti \& Piskunov~1996), 
which also synthesizes a stellar spectrum
under the assumption of LTE in a 1-D stellar atmospheric model. One can
then optimize the stellar model parameters (T$_{\rm eff}$, log~$g$ and [Fe/H])
by minimizing the $\chi^2$ difference between the observed and the synthesized SME
spectra.  This can be a very effective and efficient method for
deriving uniform and self-consistent stellar parameters for large samples
of spectra (e.g. Valenti \& Fischer~2005). All three of these
techniques work best on spectra of relatively high S/N and high
spectral resolution.

The analysis of lower S/N spectra requires special approaches in
order to achieve useful and self-consistent results.  When the spectra are
sufficiently noisy that individual stellar absorption lines can not be
measured reliably, then one must adopt a technique that will treat large
sections of the spectrum together.  The most successful of these methods
involve some form of spectral matching, where the observed spectrum is
compared to a library of spectra, and a comparison algorithm determines the
``best'' match to the observations.  In some cases, the library is a set of
observed spectra with carefully and self-consistently determined stellar
properties, as in the {\tt SpecMatch} code used by Petigura, Marcy, \& Howard~(2013).
For this to work well, the library of
observed spectra must be obtained with the same instrument as the target
spectrum, they must all be of high signal/noise, and they need to cover the
parameter space of T$_{\rm eff}$, log~$g$ and [Fe/H] rather uniformly.
A viable alternative to this technique is use a library of synthetic model
spectra, rather than observed spectra.  This ensures that the relevant
ranges of parameter space are well covered, and that all of the library
spectra are completely self-consistent and noise-free. However, one must
model the spectrograph instrumental function in order to compare an
observed spectrum with the library spectra.  This approach of synthetic
spectra has been used by Buchhave et al.~(2012) to perform a uniform analysis of relatively low signal/noise
spectra from different spectrographs.   A new version of {\tt SpecMatch}
now uses a grid of synthetic models (Petigura et al. 2015, in preparation).

In this paper, we present {\it Kea}, a code that we developed at McDonald
Observatory to compare high-resolution, low S/N spectra of KOI stars
to a massive grid of synthetic stellar
spectral models in order to determine the fundamental stellar parameters of
the {\it Kepler} target stars.
Our paper is organized as follows: section 2 describes the our spectroscopic observations and data reduction. In Section 3, we detail how we created a grid of 
synthetic spectra using the MOOG spectrum synthesizer. In Section 4, we describes the calibration of {\it Kea} using
100 well-characterized stars from the {\it Kepler} stellar properties catalog (Huber et al.~2014, = H14 hereafter), the so-called ``platinum'' star sample.
Finally, in Section 5 we present a comparison of {\it Kea} results from our McDonald Observatory reconnaissance spectra with stellar parameters derived from 
higher S/N Keck/HIRES spectra of the same KOI.

\section{Reconnaissance Spectroscopy \& Data Reduction}

We use the Tull Coud\'e Spectrograph (Tull et al.~1995) at the Harlan J. Smith 2.7\,m Telescope at McDonald Observatory to obtain the reconnaissance 
spectra. We observe with a 1.2-arcsecond slit, which yields a spectral resolving power of $R=\lambda/\delta\lambda=60,000$. The complete visual spectrum
(3750 -- 10200\,\AA) is imaged on a 2k$\times$2k CCD detector.     

After flat-fielding, bias-subtraction and order extraction, using standard IRAF routines, we divide each order by the appropriate blaze-function. We 
determine the shape for the blaze-function for each night using high-S/N flat field lamp exposures. This division removes the large-scale curvature 
due to the blaze. We then apply an additional correction to each order to remove any residual curvature in the continuum. 

The uncertainty of each pixel $\sigma_{\rm pixel}$ in the extracted spectrum is calculated as:
\begin{equation}   
\sigma_{\rm pixel} = \sqrt{N_{\rm pixel} + n\sigma_{\rm readout}^{2}}
\end{equation}
Where $N_{\rm pixel}$ are the total number of detected photo-electrons in a pixel, $n$ is the number of pixels in a column that were combined during order extraction,
and $\sigma_{\rm readout}$ is the readout noise. For the Tull spectra we use $n=5$ and $\sigma_{\rm readout} = 3.06$~electrons. 

\begin{figure}[t]
\includegraphics[angle=270,scale=0.6]{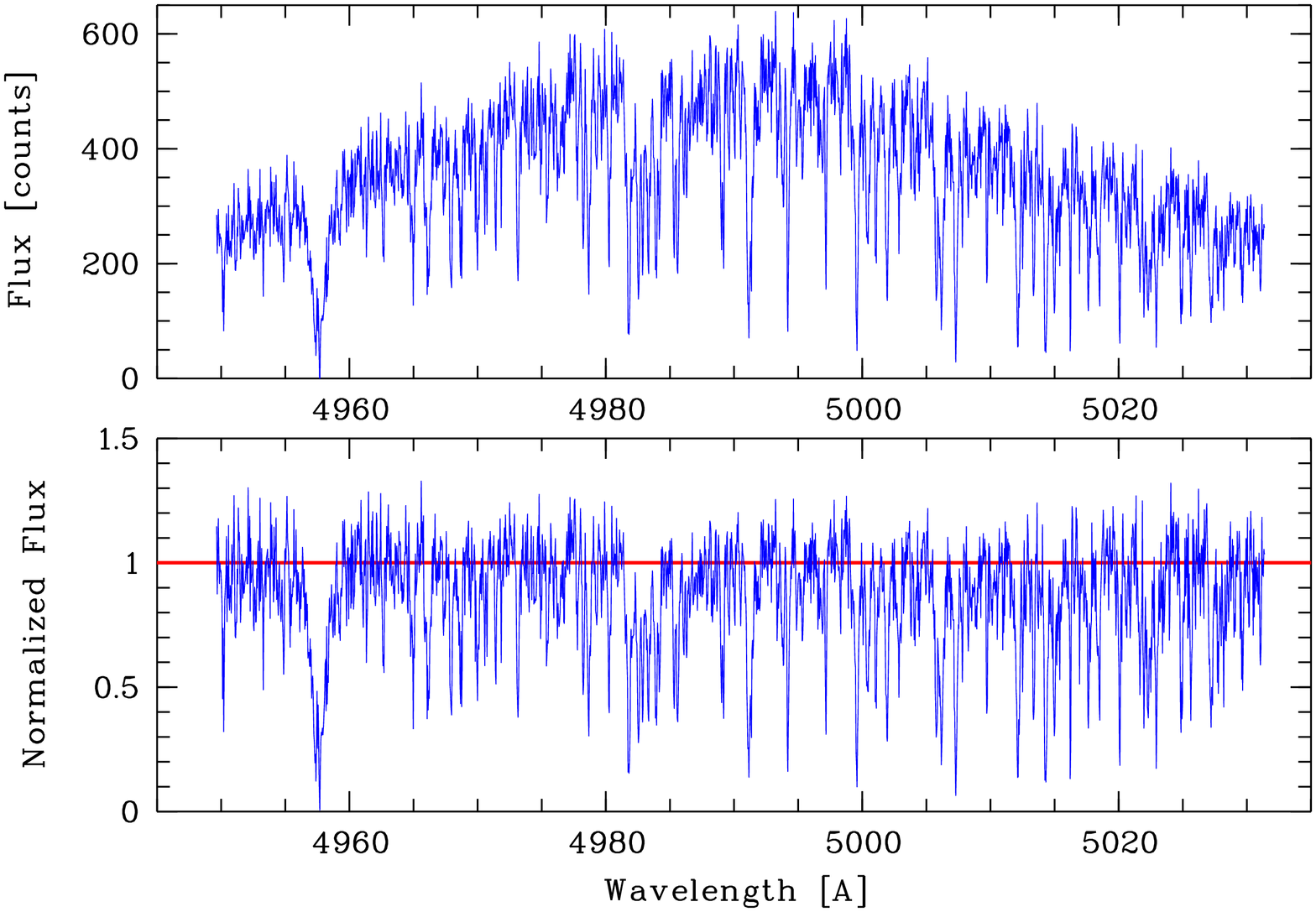}
\caption{Example of a KOI reconnaissance spectrum before (top panel) blaze division and residual continuum flattening and after (bottom panel).
The S/N of this spectrum at the top of the blaze of the order is S/N$\approx17:1$.
\label{fig1}}
\end{figure}

Finally, we flux-normalize the spectral orders to unity, scaling the pixel uncertainties accordingly. Figure\,\ref{fig1} shows an 
example of one order of a typical KOI reconnaissance spectrum before and after these preparatory steps.      

\section{Synthetic Spectral Library}

We computed a large grid of model stellar spectra using the `synth' mode of
the LTE stellar spectral line analysis and spectrum synthesis MOOG.  
We used the Kurucz~(1993) stellar atmosphere grid, with the
``ODFNEW'' opacity distribution function.   Spectra were synthesized
from 3450\,{\AA} to 7000\,{\AA}.  The complete spectral grid covers a range
of $T_{\rm eff}$ from 3500~K to 7000~K in 100~K steps, and from 7000~K
to 10,000~K in steps of 200~K.
We used [Fe/H] to represent overal stellar metallicity.
Model spectra were computed with [Fe/H] ranging from -1.0 to +0.5\,dex in
0.25\,dex steps.  All models used a solar value of [$\alpha$/Fe].
Stellar surface gravity was varied from 1.0 to 4.0 in
steps of 0.5\,dex and from 4.0 to 5.0 in steps of 0.25\,dex.  No spectra
were computed with $\log g = 1.50$ for $T_{\rm eff}$ from 9200~K to 10,000~K
nor for $\log g = 1.00$ for $T_{\rm eff}$ from 8400~K to 10,000~K as those
regions of parameter space were not covered by the Kurucz~(1993) model
atmospheres.  The final grid comprises a total of 8752 synthetic spectra.

We obtained atomic line parameters 
($\log gf$, excitation potential, and damping parameters) from
VALD (Vienna Atomic Line Database, Kupka et al.~2000).  We included
molecular opacities for MgH (Bernath et al.~1985, Hinkle et al.~2013), TiO (Plez, 1998), and CN
(Sneden et al.~2014).  The MgH line list included $^{24}$MgH lines in the 
A $^{2}${$\Pi$}-X $^{2}${$\Sigma$}$^{+}$ system listed by Bernath et al.~(1985),
supplemented by $^{25}$MgH and $^{26}$MgH lines from the compillation of
Hinkle et al.~(2013).  For CN, we limited our consideration to $^{12}$C$^{14}$N
lines in the A~$^{2}${$\Pi$}-X $^{2}${$\Sigma$}$^{+}$ (red) and 
B~$^{2}${$\Sigma$}$^{+}$-X $^{2}${$\Sigma$}$^{+}$ (violet) systems.  Both
of these linelists are available from the MOOG labdata web
page\footnote{http://www.as.utexas.edu/$\sim$chris/lab.html}.  The TiO line list
included only the $^{48}$Ti$^{16}$O isotopologue.  In all, our final line
list included approximately 3.3 million spectral lines.
For each synthetic spectrum calculation, we employed the MOOG ``weedout''
feature to remove atomic and molecular lines from the linelist with a
ratio of line to continuum opacity of less than 0.001.
A separate line list of ``strong'' spectral lines was used so that the
extended damping wings of H~Balmer lines, and certain lines of Na, Mg,
Ca, Cr, Mn and Fe could be computed fully.

\section{Fitting the Data}

Before {\it Kea} can match the synthetic spectra to the observed spectral orders, the synthetic spectra need to be convolved with the appropriate 
point-spread-function (PSF) to assure the same spectral resolution of model and data. For this purpose, we convolve the synthetic spectrum with
a Gaussian-shaped PSF and down-sample the model to set it on the same pixel scale as the observation. 
For each order we calculate the correct width of the PSF for an $R=60,000$ spectrum with 2048 model pixels. In addition to
this PSF-convolution, {\it Kea} also applies a standard rotational broadening function for stellar lines as derived in Gray~(2005). 
The rotational broadening will likely also absorb any residual PSF broadening for spectra where the resolution is slightly different to $R=60,000$.  
We did not
include macroturbulence as a line broadening effect. We think that the inclusion of macroturbulence as an additional model parameter is not warranted, given 
the typically moderate to low S/N values of the spectra that {\it Kea} is applied to. The rotational broadening will likely absorb any macroturbulence effects and 
might therefore be slightly overestimated. After these steps, the model spectrum is ready to be compared to the data. 
{\it Kea} is using the standard $\chi^{2}$ criterion for the goodness-of-fit test.         

In the next step we determine the wavelength shift $\delta\lambda$ between the model and the observation. This $\delta\lambda$ is caused by the 
combination of the absolute radial velocity of the target star and the Earth's motion at the time of observation. 
For this purpose we use one spectral order and shift a default model 
with no rotational broadening and solar $T_{\rm eff}$, [Fe/H] and log~$g$ until a $\chi^{2}$-minimum is found that corresponds to the $\delta\lambda$ 
between model and data. 
%Figure~\ref{fig2} shows an example of $\chi^{2}$ as a function of radial velocity. 
We apply the $\delta\lambda$ that corresponds to
the $\chi^{2}$-minimum to all {\it Kea} model orders for this particular spectrum. In some cases, particularly for very low S/N data, one order is not enough
to determine the $\delta\lambda$ shift. Under these circumstances we typically use 4 to 5 different spectral orders to find the correct $\delta\lambda$. 

To save computational time, {\it Kea} does not compare the entire grid of synthetic spectra to every single observed spectral order. We adopted a two-step 
approach:

1. {\it Kea} is run using a coarse step size in all four parameters spanning the entire range of the synthetic grid. From 3500~K to 7000~K $T_{\rm eff}$ we use a 
step of 500~K in $T_{\rm eff}$ and from 7000~K to 10,000~K a step of 1000~K, for [Fe/H] we use the four values of -1.0,-0.5,0.0, and +0.5, and for log~$g$ we use 
the values 1.0, 2.0, 3.0, 4.0, 4.5,and 5.0. For $v \sin i$ in the range from 1 to 15\,km\,s$^{-1}$ we move in steps of 2\,km\,s$^{-1}$ and then from 
20 to 60\,km\,s$^{-1}$ in steps of 10\,km\,s$^{-1}$. For each spectral order, {\it Kea} determines the
$\chi^{2}$-value and records the best-fit model for each order. The final parameters and uncertainties that {\it Kea} reports for a spectrum are the mean values 
of the model parameters of all best-fit models and the formal uncertainty of this mean (=RMS/$\sqrt{N}$ with $N$ the number of orders used to determine the 
mean). This initial run yields a first ``guess'' set of stellar parameters that we use as input parameters for step 2. In case of a fast rotator with $v \sin i > 
60$\,km\,s$^{-1}$ the best-fit models for all orders will have the maximum value of 60\,km\,s$^{-1}$. For these targets we expand the range of trial $v \sin i$ 
values even further.

2. in the second step, we run {\it Kea} using the densest possible step size that is set by the model grid itself: $T_{\rm eff}$ in 100~K, [Fe/H] and log~$g$ in 
0.25 dex steps, and $v \sin i$ in 1\,km\,s$^{-1}$ steps (except for fast rotators where we use steps of 5-10\,km\,s$^{-1}$). 
In contrast to the previous step we now limit 
the range that {\it Kea} searches in the model grid to $T_{\rm eff}$ values that are $\pm500$~K, $\pm1.5$dex in log~$g$ and $\pm5$\,km\,s$^{-1}$ in $v \sin i$ 
from the first-guess values from step 1 using the whole range of [Fe/H] values in our library. 
As before, we record the best-fit ($\chi^{2}$-minimum) models for each spectral order that {\it Kea} analyzes and determine the final stellar parameters from 
the mean and scatter of these values. 

\begin{figure}
\includegraphics[angle=270,scale=0.62]{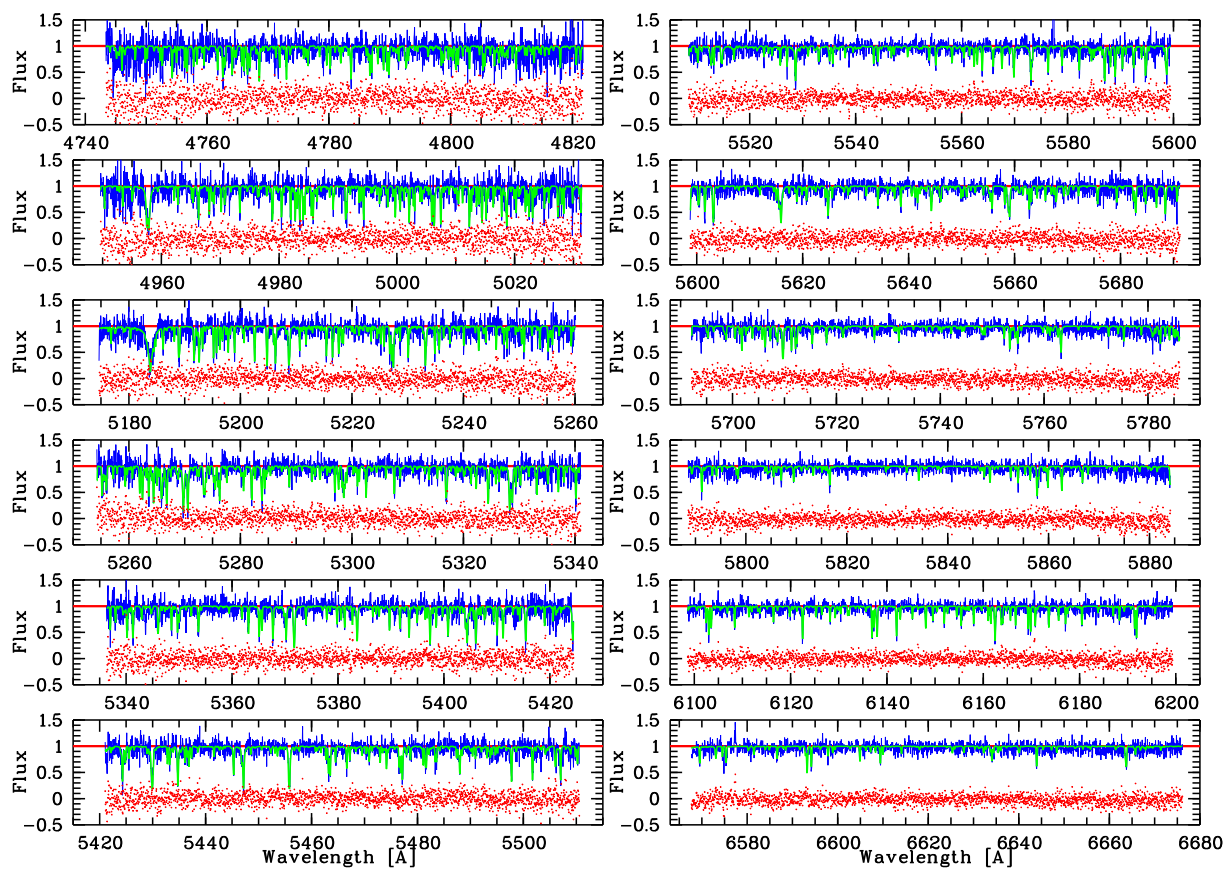}
\caption{Comparison of an observed spectrum (blue) to best-fit {\it Kea} models in green (residuals in red) for 12 spectral orders of one
typical KOI reconnaisssance spectrum with a S/N of 23 at 5650~\AA.
\label{fig3}}
\end{figure}

Figure\,\ref{fig3} shows 12 spectral orders from one observed KOI spectrum (blue) and the corresponding best-fit {\it Kea} model in green and the residuals in 
red.    

\section{Calibration with Platinum star sample}

To test {\it Kea} we used the so-called ``platinum'' star sample of the {\it Kepler} follow-up observing program. The platinum
stars are a carefully selected group of stars from the {\it Kepler} stellar properties catalog (H14) that all have 
asteroseismically derived log~$g$ values with very small uncertainties of the order of $0.03$ dex. 
During the 2014 {\it Kepler} observing season we collected 
spectra for 100 platinum stars, using the exact same instrumental setup as for the KOI observations. The stars range in $T_{\rm eff}$ from 5000 to 6700~K, in 
[Fe/H] from -1.0 to +0.5 and in log~$g$ from 3.3 to 4.6 dex. The majority of the sample are slow rotators with only 12 stars having a $v \sin i > 10$\,km\,s$^{-1}$.
Out of these 12 only 5 stars have $v \sin i > 30$\,km\,s$^{-1}$. A detailed description of the selection criteria of the platinum sample and a
comparison of different methods to derive stellar parameters will be presented in a future publication (Furlan et al.~2016, in prep.)  

The platinum stars are also significantly brighter than the KOIs we observe at 
McDonald Observatory, which reflects in a much higher S/N for the platinum star spectra. Typically, a KOI reconnaissance spectrum has a S/N of 20-30 per 
resolution element at 5650 {\AA}, while the platinum star spectra have S/N$\approx80$.   

\begin{figure}
\includegraphics[angle=0,scale=0.68]{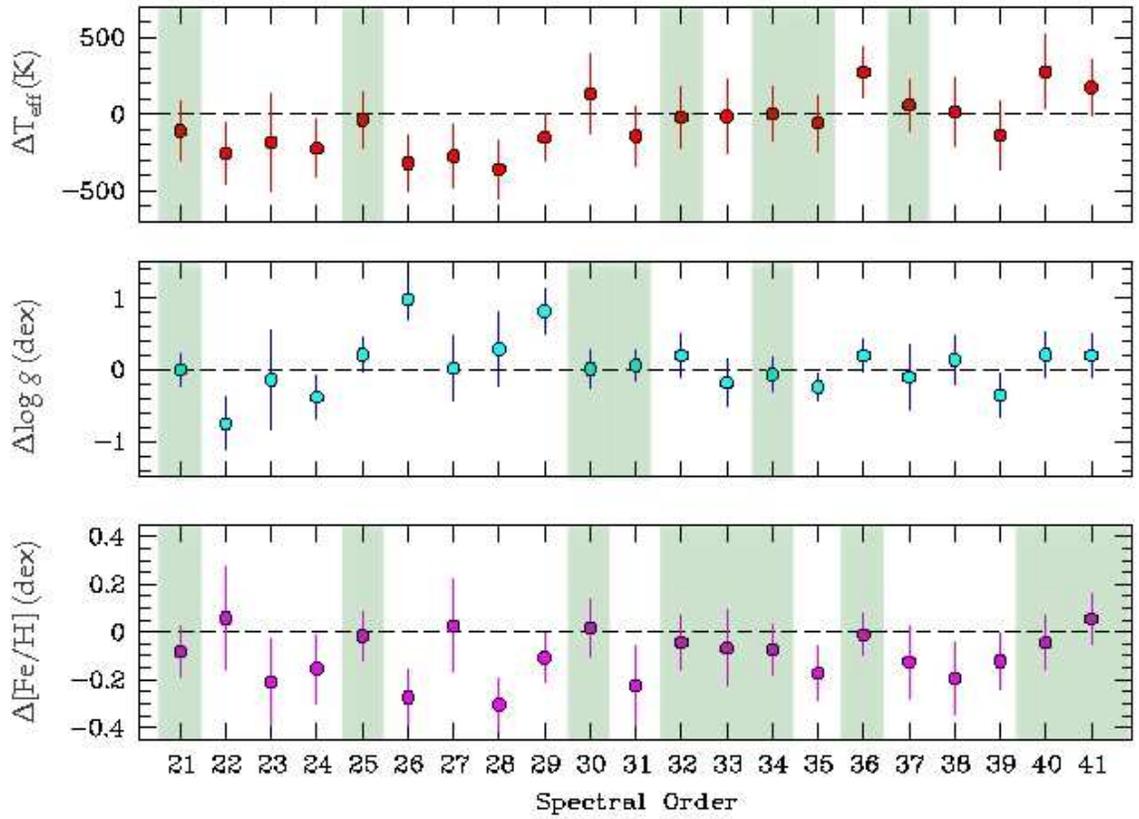}
\caption{The difference between individual spectral order results from {\it Kea} and the published values from
Huber et al.~(2014). The 13 orders that are shown with a green shaded background were selected to provide information
on the particular stellar parameter (see text for details on the selection criteria).   
\label{orders}}
\end{figure}

We used {\it Kea} to derive stellar parameters from these 100 spectra. We compared the overall mean offset and RMS-scatter of the 100 {\it Kea} values with 
the published values for each of the 21 orders of the Tull spectrum that covers the wavelength range of our library. Table\,\ref{tab1} contains the
complete information of this order by order comparison and the result is displayed in 
Figure\,\ref{orders}. We calibrateb {\it Kea} by testing which spectral orders yield 
the smallest offsets from, and smallest scatter around, the reported values in H14. 
With this procedure we identify spectral orders that
are sensitive to the stellar parameter that we want to determine. We achieved best results by using 13 (out of these 21) spectral 
orders which satisfy the following criteria: the mean offset in $T_{\rm eff}$ is less than 110~K and the overall scatter of the
values is less than 200K, the offset in [Fe/H] is less then 0.1~dex and the RMS is less than 0.2~dex, and for log~$g$ we selected
orders that have an offset of less than 0.1~dex and an RMS less than 0.3~dex. The resulting selection of orders are displayed in
Figure\,\ref{orders} with a (green) shaded background.   
%summarized in Table\,\ref{tab1}. 
%We also indicate in the table which orders {\it Kea} is using to obtain information on a particular stellar 
%parameter like $T_{\rm eff}$, [Fe/H] or log~$g$. 
All 13 orders are being used to measure the $v \sin i$.  

%Figures\,\ref{huber1} to \ref{huber3} show the comparison between the {\it Kea} results and the reported stellar parameters from Huber et al.~(2014) for our 
%sample of 
%platinum stars. 
For $T_{\rm eff}$ and [Fe/H] we can perform this calibration only for a subset of 63 stars that also have spectroscopically derived [Fe/H] and 
$T_{\rm eff}$ listed in H14 (the remaining 36 stars have these values estimated from photometry and [Fe/H] set to a default value of 
$-0.2\pm0.3$). The uncertainties in $T_{\rm eff}$ for these 63 stars is $119\pm8$~K. The uncertainties for the spectroscopically derived [Fe/H] values
in H14 are given as 0.15~dex.  
Our calibration yield in effective temperature a small offset of -25~K with an RMS-scatter of 89~K. For the metallicity we achieved an offset of -0.02 dex 
and an RMS of 0.07 dex. For surface gravity we can use all 100 stars and compare it to the asteroseismically derived log~$g$ values. The {\it Kea} results show a small
offset of -0.0006 dex, smaller than the H14 uncertainties, and an RMS-scatter of 0.11~dex. 
   
\begin{figure}
\includegraphics[angle=270,scale=0.62]{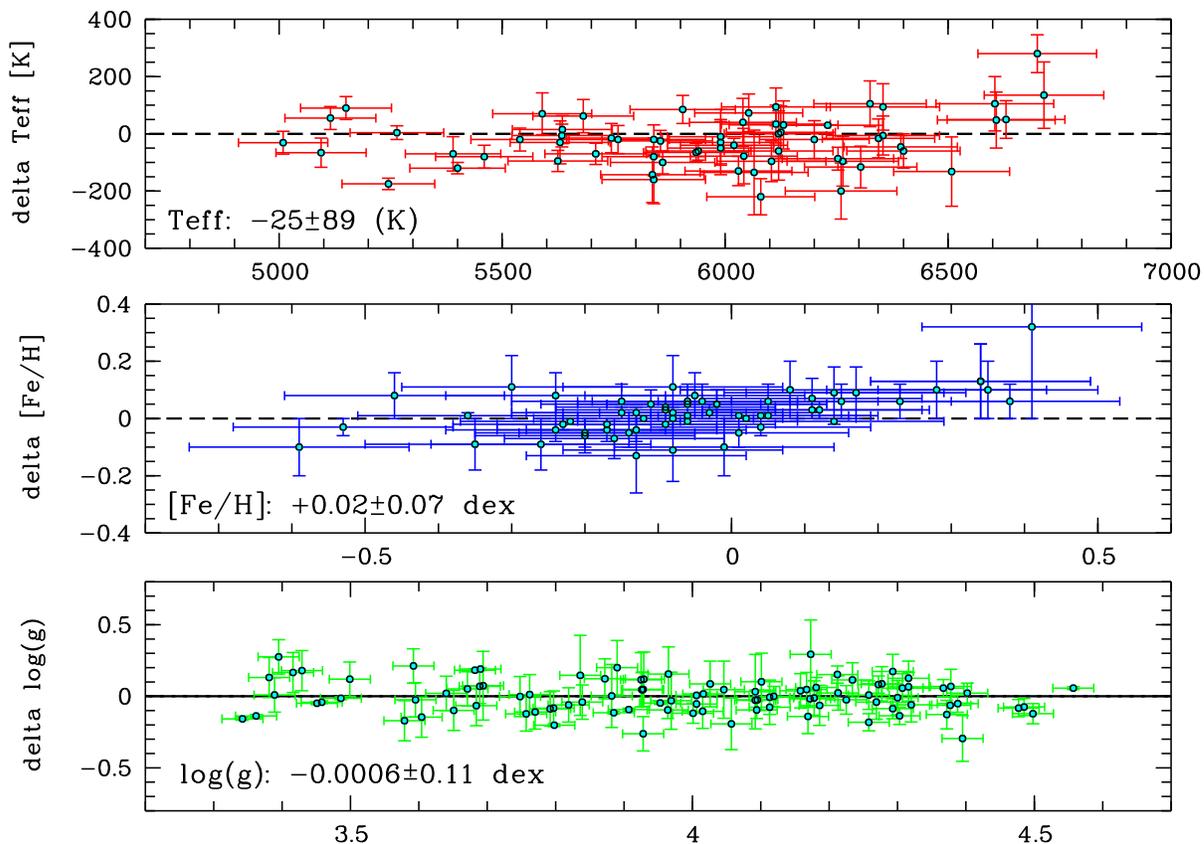}
\caption{The difference between {\it Kea} results and the Huber et al.~(2014) catalog as function of 
the effective temperature of the star (upper panel), metallicity (middle panel) and log~$g$ (bottom panel).
There are no obvious systematic trends visible. In effective temperature we have an offset of -25 K and an RMS-scatter 
of 89 K. In [Fe/H] we find a mean offset of +0.02 with an RMS-scatter of 0.07 dex. In log\,$g$ the respective values are
-0.0006 and 0.11~dex.
\label{huber4}}
\end{figure}

In Figure\,\ref{huber4} we display the dependence of the difference between the {\it Kea} results and the values from H14 on the value of this 
parameter. We do not see any strong systematic trends in these differences.  

\section{Comparison with Keck/HIRES SME results}

After the calibration with the high S/N platinum star spectra, we tested {\it Kea} on data in the S/N range that is typical for our
{\it Kepler} mission reconnaissance observations. 
We compared the {\it Kea} results for 45 reconnaissance spectra of 32 KOIs from the beginning of the mission (all with KOI numbers $<1000$) with the results 
derived from Keck/HIRES spectra. The Keck data were analyzed with SME and we took the stellar parameters that are posted in the notes section of 
each KOI on the CFOP webpage\footnote{https://cfop.ipac.caltech.edu/}. In a few cases, we took the stellar parameters from the published literature: e.g. KOI-87: 
Borucki et al.~(2012); KOI-128: Endl et al.~(2012); KOI-135, KOI-183 and KOI-214: Endl et al.~(2014). The S/N of the 45 Tull reconnaissance spectra at 
5650 {\AA} ranges from 13 to 133 per resolution element, with a mean of 28 and a standard deviation of 19. The effective temperature range of these
data is 4700 to 6100~K, in [Fe/H] from -0.55 to +0.45 and in log $g$ from 3.9 to 4.7~dex.   

\begin{figure}
\includegraphics[angle=270,scale=0.62]{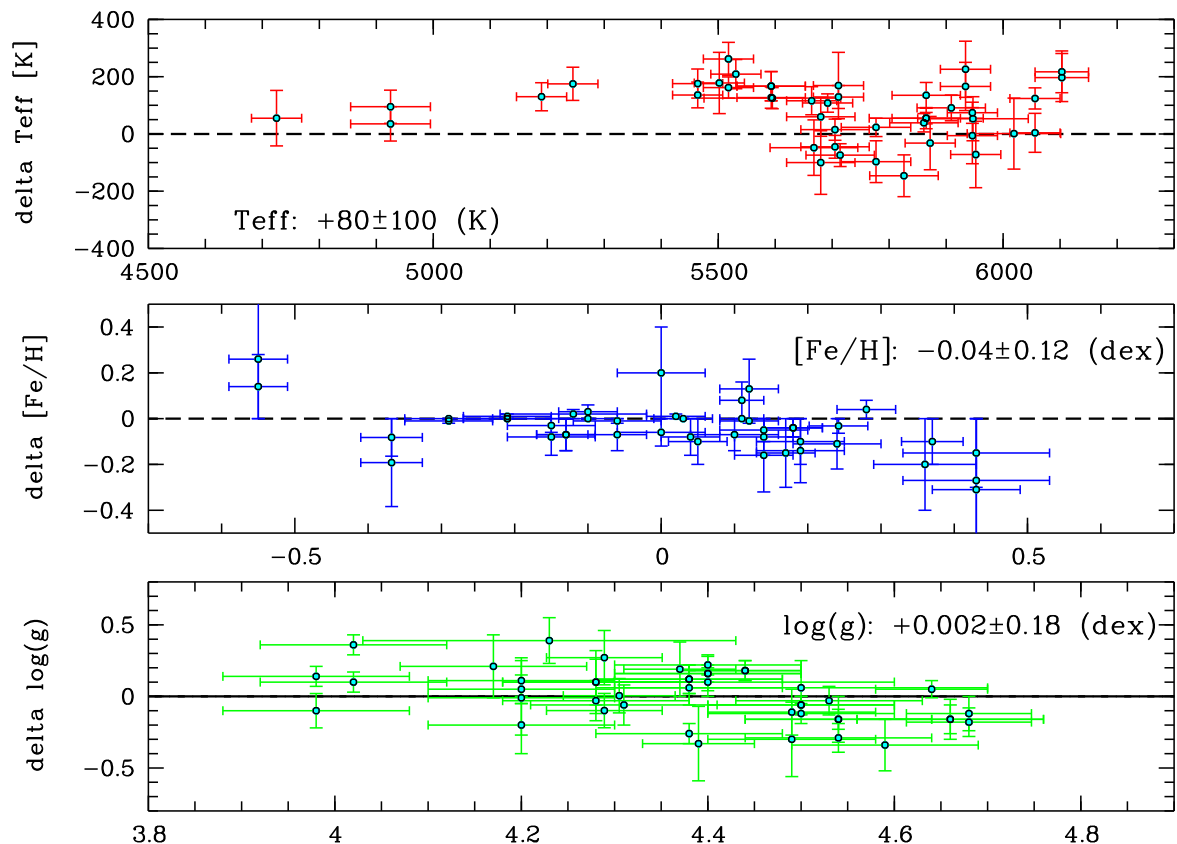}
\caption{Comparison of the {\it Kea} results for 45 reconnaissance spectra with the parameters obtained with SME from
Keck/HIRES spectra (posted on CFOP website). In $T_{\rm eff}$ we find an offset of 80~K (the Kea values are higher) and an RMS-scatter of 100~K, 
in [Fe/H] the offset is -0.04 with an RMS of 0.12 dex and in log~$g$ the offset is +0.002 with an RMS of 0.18 dex. 
\label{keck1}}
\end{figure}

We display the results in Figure\,\ref{keck1}. In effective temperature we find an average offset
of +80 K and an RMS-scatter of 100 K. The {\it Kea} $T_{\rm eff}$ values are systematically higher than the SME values, especially in the 5200 to 5600 K range.
For the metallicity parameter we see an offset of-0.04 dex and a standard deviation of 0.12 dex. And for the surface gravity 
we find a very small offset of +0.002 dex with an RMS-scatter of 0.18 dex between the Keck results and ours. 
Table\,\ref{tab2} lists the SME and {\it Kea} values from this test along with the KOI number and the S/N of the Tull spectra (we do not have access to the
S/N values of the HIRES spectra). 

The slightly larger offset in $T_{\rm eff}$ and the increased scatter of these results, as compared to the
platinum star sample calibration, might be due to the lower S/N of the Tull spectra. We tested this hypothesis by artifically degrading the S/N of the
platinum star spectra. We used a subset of 30 spectra that originally have $\Delta T_{\rm eff}=-24\pm110~K$, $\Delta$[Fe/H] = $0.02\pm0.09$~dex and
$\Delta$~log~$g=0.0006\pm0.108$~dex. Degrading these spectra to S/N=30 yielded the following values: $\Delta T_{\rm eff}=13\pm123~K$, $\Delta$[Fe/H] = $0.06\pm0.09$~dex 
and $\Delta$~log~$g=0.04\pm0.19$~dex. Decreasing the S/N further down to S/N=20 we obtained $\Delta T_{\rm eff}=33\pm124~K$, $\Delta$[Fe/H] = $0.08\pm0.11$~dex and
$\Delta$~log~$g=0.22\pm0.29$~dex. These results indicate that a major contribution to the larger offsets and RMS-scatter for the SME-{\it Kea} comparison is simply 
lower S/N.  
%If we just compare the 23 spectra with S/N$>30$ the RMS-scatter decreases to 79~K in 
%$T_{\rm eff}$, 0.075 dex in [Fe/H] and to 0.18 dex in log~$g$, which is fully consistent to the platinum star results. 
%In Figure\,\ref{keck2} we show the differences between the {\it Kea} results and the Keck results as function of S/N of the Tull reconnaissance spectra. Except 
%for the 3 lowest S/N spectra, we find no apparent systematic trends in these data.  

\section{Limitations of {\it Kea}}

Owing to the specific calibration of {\it Kea} we note that results for stars with effective temperatures outside the range of the platinum sample (5000 - 6700~K) 
might have larger uncertainties than the ones quoted here. Also, {\it Kea} is not tested for rapid rotators with $v \sin i > 30$\,km\,s$^{-1}$. Finally, our
results indicate that stellar parameters derived with {\it Kea} from spectra with S/N less than 20 are unreliable. 

\section{Conclusions}

We present a description of the new {\it Kea} spectroscopic fitting tool, which we use to derive stellar parameters ($T_{\rm eff}$, [Fe/H], log~$g$ and $v \sin 
i$) for the {\it Kepler} mission reconnaissance spectra that we collect with the Tull spectrograph at the Harlan J. Smith 2.7\,m telescope at McDonald Observatory. 
A calibration with the sample of 100 platinum stars yield typical uncertainties of $\pm 90$~K in effective temperature, of $\pm 0.07$ dex in [Fe/H] and 
of $\pm 0.11$ dex in log~$g$. We tested {\it Kea} by comparing it to stellar parameters derived from higher S/N Keck/HIRES spectra for 32 KOIs and 45 
Tull spectra in the S/N range of our reconnaissance observations. We find a typical RMS-scatter of 100~K in $T_{\rm eff}$, 0.12 dex in [Fe/H], and 
0.18 dex in log~$g$. 
    
{\it Kea} is now in routine operation to obtain spectroscopically determined parameters from the KOI reconnaissance data that the McDonald Observatory {\it 
Kepler} follow-up observing team are collecting. We have used {\it Kea} recently to analyze spectra of Kepler-452 (Jenkins et al.~2015), a G2 star orbited 
by a 1.6~R$_{\oplus}$ planet inside its circumstellar habitable zone.

\acknowledgments
The {\it Kepler} Follow-Up Observing Program observations at McDonald Observatory
are supported by NASA Grant NNX13AB62A from NASA Ames Research Center. ME acknowledges support by NASA under Grant NNX14AB86G issued through the Kepler
Participating Scientist Program. This work has made use of the VALD database, operated at Uppsala
University, the Institute of Astronomy RAS in Moscow, and the University of
Vienna.  We are extremely grateful to Chris Sneden and Erik Brugamyer for
their patient advice and assistance in our use of MOOG, and to Phillip J. MacQueen, who helped with many important discussions during
the development of {\it Kea}. We also thank the entire {\it Kepler} community follow-up observing (CFOP) working group, led by David Ciardi, who
helped with coordianting the reconnaissance and platinum star observations. 
Funding for the Discovery mission {\it Kepler} is provided by NASA's Science Mission 
Directorate.

%\begin{figure}
%\includegraphics[angle=270,scale=0.6]{keapap_fig2.eps}
%\caption{Example of one determination of the radial velocity shift ($\delta\lambda$) between the {\it Kea} model and the observed spectrum.
%In this case, a radial velocity shift of -38.5\,km\,s$^{-1}$ was determined. 
%\label{fig2}}
%\end{figure}

%\begin{figure}
%\includegraphics[angle=270,scale=0.62]{keapap_fig4.eps}
%\caption{Comparison of the reported $T_{\rm eff}$ values from Huber et al.~(2014) with the {\it Kea} results for 63 platinum
%stars. We find a small offset of -29 K and an RMS-scatter of 98 K. The mean formal uncertainty of the
%{\it Kea} values is 82~K.
%\label{huber1}}
%\end{figure}

%\begin{figure}
%\includegraphics[angle=270,scale=0.62]{keapap_fig5.eps}
%\caption{Comparison of the reported [Fe/H] values from Huber et al.~(2014) with the {\it Kea} results for 63 platinum
%stars. We find a small offset of -0.04 dex and an RMS-scatter of 0.093 dex. The mean formal uncertainty of the
%{\it Kea} values is 0.06 dex.
%\label{huber2}}
%\end{figure}

%\begin{figure}
%\includegraphics[angle=270,scale=0.62]{keapap_fig6.eps}
%\caption{Comparison of the reported log~$g$ values from Huber et al.~(2014) with the {\it Kea} results for 100 platinum
%stars. We find a small offset of +0.023 dex and an RMS-scatter of 0.16 dex. The mean formal uncertainty of the
%{\it Kea} values is 0.12 dex.
%\label{huber3}}
%\end{figure}

%\begin{figure}
%\includegraphics[angle=270,scale=0.62]{sme_comp_panel_SNR.eps}
%\caption{The difference between {\it Kea} results for 46 reconnaissance spectra with the parameters derived from
%Keck/HIRES spectra as function of the S/N of the Tull spectra. Just the 3 lowest S/N spectra (S/N $< 18$) show significant systematic
%differences between these two data sets.  
%\label{keck2}}
%\end{figure}

\begin{deluxetable}{ccrrr}
\tablecolumns{5}
\tablewidth{0pt}
\tablecaption{The comparison of the {\it Kea} results from all Tull spectral orders with the Huber et al.~(2014) values. 
In each column we list the average offset and RMS-scatter of the {\it Kea} values.
A "$\star$" indicates that this order is selected to obtain information on this particular parameter (see also 
Figure\,\ref{orders}).  
\label{tab1}}
\tablehead{
\colhead{order} & {$\lambda$ range} & {$\Delta T_{\rm eff}$} & {$\Delta$ [Fe/H]} & {$\Delta$ log~$g$}\\
{} & {({\AA})} &  {(K)} & {(dex)} & {(dex)}  
}
\startdata
41 & 4740 - 4820 & $173.4\pm177 $  & $0.05\pm0.11\,\star$ &  $0.20\pm0.31$   \\
40 & 4810 - 4890 &  $275.3\pm242 $ & $-0.05\pm0.11\,\star $ &  $0.21\pm0.31 $  \\
39 & 4880 - 4960 &  $-135.6\pm223 $ & $-0.12\pm0.12 $ &  $-0.35\pm0.31$  \\
38 & 4940 - 5030 &  $14.4\pm224 $ & $-0.19\pm0.15 $ & $0.14\pm0.34 $ \\
37 & 5020 - 5110 & $59.4\pm171\,\star$ & $-0.13\pm0.15 $ &  $-0.10\pm0.46 $  \\
36 & 5090 - 5180 & $272.4\pm166 $ & $-0.01\pm0.09\,\star$  & $0.20\pm0.23 $   \\
35 & 5180 - 5260 & $-58.6\pm182\,\star$ & $-0.17\pm0.11 $ & $-0.24\pm0.20$   \\
34 & 5260 - 5340 & $0.4\pm173\,\star$ & $-0.07\pm0.10\,\star$ & $-0.07\pm0.24\,\star$ \\
33 & 5330 - 5420 & $-15.6\pm243$ & $-0.07\pm0.16\,\star$ & $-0.18\pm0.33$ \\
32 & 5420 - 5520 & $-19.6\pm195\,\star$ & $-0.04\pm0.11\,\star$ &  $0.20\pm0.31$  \\
31 & 5500 - 5600 & $-142.6\pm195$ & $-0.22\pm0.17$ & $0.06\pm0.22\,\star$ \\
30 & 5600 - 5690 & $132.4\pm253$ & $0.02\pm0.12\,\star$ & $0.01\pm0.26\,\star$\\
29 & 5690 - 5790 & $-152.6\pm152$ & $-0.11\pm0.10$ & $0.81\pm0.31$  \\
28 & 5780 - 5890 & $-359.6\pm187$ & $-0.30\pm0.11$ & $0.29\pm0.52$ \\
27 & 5890 - 5985 & $-275.6\pm206$ & $0.03\pm0.19\,\star$ & $0.02\pm0.46$ \\
26 & 5990 - 6090 & $-318.6\pm185$ & $-0.27\pm0.12$ & $0.98\pm0.29$ \\
25 & 6100 - 6200 & $-37.6\pm177\,\star$ & $-0.02\pm0.10\,\star$ & $0.21\pm0.24$ \\
24 & 6210 - 6315 & $-224.6\pm190$ & $-0.15\pm0.14$ & $-0.38\pm0.30$ \\
23 & 6335 - 6430 & $-183.6\pm319$ & $-0.21\pm0.18$ & $-0.14\pm0.69$ \\
22 & 6440 - 6550 & $-258.6\pm198$ & $0.06\pm0.21$ & $-0.75\pm0.37$ \\
21 & 6550 - 6680 & $-109.6\pm190\,\star$ & $-0.08\pm0.105\,\star$ & $-0.001\pm0.24\,\star$  \\
\enddata
\end{deluxetable}

\newpage
\begin{deluxetable}{crcrrccccrrccc}
\tablecolumns{14}
\tablewidth{0pt}
\tablecaption{The comparison of the {\it Kea} results for 45 KOIs with
the values from an SME analysis of Keck/HIRES spectra of the same object.
\label{tab2}}
\tablehead{
\colhead{KOI} & {S/N} & {$T_{\rm eff}$} & {$\sigma$} & {{\small [Fe/H]}} & {$\sigma$} & 
{log~$g$} & {$\sigma$} & {$T_{\rm eff}$} & {$\sigma$} & {{\small [Fe/H]}} & {$\sigma$} & 
{log~$g$} & {$\sigma$}\\
{} & {5650\AA} & {{\small SME}} & {{\small SME}} & {{\small SME}} & {{\small SME}} &
{{\small SME}} & {{\small SME}} & {{\it Kea}} & {{\it Kea}} & {{\it Kea}} & {{\it Kea}} &
{{\it Kea}} & {{\it Kea}}
}
\startdata
005 & 43.8 & 5861. &  60. &  0.14 & 0.06 &  4.17 & 0.10 & 5900. &  32. &  0.09 & 0.03 &  4.38 & 0.22 \\
007 & 44.3 & 5865. &  60. &  0.19 & 0.06 &  4.28 & 0.10 & 6000. &  45. &  0.05 & 0.04 &  4.38 & 0.22 \\
007 & 66.5 & 5865. &  60. &  0.19 & 0.06 &  4.28 & 0.10 & 5920. &  20. &  0.09 & 0.03 &  4.25 & 0.14 \\
017 & 29.0 & 5826. &  60. &  0.43 & 0.06 &  4.59 & 0.10 & 5680. &  73. &  0.12 & 0.05 &  4.25 & 0.18 \\
041 & 43.3 & 5909. &  60. &  0.10 & 0.06 &  4.28 & 0.10 & 6000. &  45. &  0.03 & 0.02 &  4.38 & 0.16 \\
069 & 52.0 & 5593. &  60. & -0.21 & 0.06 &  4.50 & 0.10 & 5760. &  51. & -0.21 & 0.02 &  4.44 & 0.06 \\
069 & 53.3 & 5593. &  60. & -0.21 & 0.06 &  4.50 & 0.10 & 5760. &  51. & -0.20 & 0.03 &  4.38 & 0.07 \\
069 & 80.3 & 5593. &  60. & -0.21 & 0.06 &  4.50 & 0.10 & 5720. &  37. & -0.21 & 0.02 &  4.44 & 0.06 \\
072 & 48.2 & 5705. &  60. & -0.15 & 0.06 &  4.54 & 0.10 & 5660. &  40. & -0.23 & 0.04 &  4.25 & 0.10 \\
072 & 133.0 & 5705. &  60. & -0.15 & 0.06 &  4.54 & 0.10 & 5720. &  37. & -0.18 & 0.02 &  4.38 & 0.07 \\
084 & 53.0 & 5464. &  44. & -0.13 & 0.04 &  4.38 & 0.10 & 5640. &  51. & -0.20 & 0.03 &  4.50 & 0.10 \\
084 & 59.3 & 5464. &  44. & -0.13 & 0.04 &  4.38 & 0.10 & 5600. &  45. & -0.20 & 0.02 &  4.50 & 0.10 \\
087 & 55.2 & 5518. &  44. & -0.29 & 0.06 &  4.44 & 0.06 & 5680. &  37. & -0.30 & 0.03 &  4.62 & 0.07 \\
087 & 59.4 & 5518. &  44. & -0.29 & 0.06 &  4.44 & 0.06 & 5780. &  58. & -0.29 & 0.02 &  4.62 & 0.07 \\
097 & 30.4 & 5934. &  44. &  0.11 & 0.03 &  3.98 & 0.10 & 6160. &  98. &  0.19 & 0.03 &  3.88 & 0.12 \\
097 & 45.8 & 5934. &  44. &  0.11 & 0.03 &  3.98 & 0.10 & 6100. &  84. &  0.11 & 0.02 &  4.12 & 0.07 \\
103 & 30.3 & 5531. &  44. & -0.06 & 0.04 &  4.40 & 0.10 & 5740. &  51. & -0.13 & 0.03 &  4.62 & 0.07 \\
108 & 28.0 & 5946. &  44. &  0.14 & 0.04 &  4.20 & 0.10 & 5940. &  98. &  0.06 & 0.04 &  4.25 & 0.10 \\
108 & 49.5 & 5946. &  44. &  0.14 & 0.04 &  4.20 & 0.10 & 6020. &  37. & -0.02 & 0.04 &  4.31 & 0.16 \\
111 & 23.6 & 5711. &  44. & -0.55 & 0.04 &  4.20 & 0.10 & 5880. & 116. & -0.29 & 0.06 &  4.19 & 0.26 \\
111 & 27.0 & 5711. &  44. & -0.55 & 0.04 &  4.20 & 0.10 & 5840. &  40. & -0.41 & 0.03 &  4.00 & 0.20 \\
116 & 39.6 & 5865. & 100. & -0.12 & 0.10 &  4.40 & 0.20 & 5920. &  37. & -0.10 & 0.04 &  4.50 & 0.10 \\
122 & 37.6 & 5714. &  60. &  0.24 & 0.06 &  4.38 & 0.10 & 5640. &  40. &  0.13 & 0.03 &  4.12 & 0.07 \\
123 & 37.9 & 5947. &  97. & -0.06 & 0.06 &  4.31 & 0.10 & 6000. &  77. & -0.07 & 0.04 &  4.25 & 0.14 \\
127 & 27.7 & 5668. &  77. &  0.43 & 0.10 &  4.53 & 0.10 & 5620. &  97. &  0.16 & 0.03 &  4.50 & 0.10 \\
128 & 25.9 & 5595. & 120. &  0.36 & 0.07 &  4.23 & 0.20 & 5720. &  37. &  0.16 & 0.03 &  4.62 & 0.16 \\
135 & 27.4 & 6019. &  82. &  0.43 & 0.10 &  4.54 & 0.10 & 6020. & 124. &  0.28 & 0.04 &  4.38 & 0.16 \\
139 & 20.4 & 5952. &  44. &  0.28 & 0.04 &  4.38 & 0.10 & 5880. & 116. &  0.32 & 0.04 &  4.44 & 0.12 \\
148 & 35.4 & 5190. &  44. &  0.17 & 0.04 &  4.40 & 0.10 & 5320. &  49. &  0.02 & 0.03 &  4.56 & 0.12 \\
153 & 22.1 & 4725. &  44. &  0.05 & 0.04 &  4.50 & 0.20 & 4780. &  97. & -0.05 & 0.09 &  4.56 & 0.19 \\
180 & 19.0 & 5680. &  60. &  0.00 & 0.06 &  4.66 & 0.10 & 5580. & 111. &  0.20 & 0.04 &  4.50 & 0.14 \\
180 & 34.1 & 5680. &  60. &  0.00 & 0.06 &  4.66 & 0.10 & 5740. &  68. & -0.06 & 0.02 &  4.50 & 0.10 \\
244 & 36.8 & 6103. &  47. & -0.10 & 0.04 &  4.02 & 0.10 & 6300. &  84. & -0.07 & 0.03 &  4.38 & 0.07 \\
244 & 39.6 & 6103. &  47. & -0.10 & 0.04 &  4.02 & 0.10 & 6320. &  73. & -0.10 & 0.02 &  4.12 & 0.07 \\
261 & 37.5 & 5692. &  44. &  0.04 & 0.04 &  4.40 & 0.10 & 5800. &  32. & -0.04 & 0.03 &  4.56 & 0.06 \\
265 & 32.7 & 6056. &  44. &  0.18 & 0.04 &  4.29 & 0.06 & 6060. &  68. &  0.14 & 0.03 &  4.19 & 0.12 \\
265 & 33.7 & 6056. &  44. &  0.18 & 0.04 &  4.29 & 0.06 & 6180. &  37. &  0.14 & 0.03 &  4.56 & 0.19 \\
273 & 39.3 & 5664. &  44. &  0.24 & 0.04 &  4.30 & 0.06 & 5780. &  49. &  0.21 & 0.03 &  4.31 & 0.12 \\
372 & 21.8 & 5872. &  44. &  0.02 & 0.04 &  4.64 & 0.06 & 5840. &  93. &  0.03 & 0.04 &  4.69 & 0.06 \\
377 & 18.8 & 5777. &  61. &  0.12 & 0.04 &  4.49 & 0.09 & 5680. &  73. &  0.25 & 0.06 &  4.19 & 0.26 \\
377 & 23.2 & 5777. &  61. &  0.12 & 0.04 &  4.49 & 0.09 & 5800. &  32. &  0.11 & 0.03 &  4.38 & 0.16 \\
555 & 19.9 & 5245. &  44. &  0.03 & 0.04 &  4.37 & 0.06 & 5420. &  58. &  0.03 & 0.04 &  4.56 & 0.19 \\
701 & 22.8 & 4925. &  70. & -0.37 & 0.04 &  4.68 & 0.07 & 5020. &  58. & -0.56 & 0.04 &  4.50 & 0.10 \\
701 & 23.1 & 4925. &  70. & -0.37 & 0.04 &  4.68 & 0.07 & 4960. &  60. & -0.45 & 0.04 &  4.56 & 0.12 \\
711 & 19.2 & 5502. &  44. &  0.37 & 0.04 &  4.39 & 0.06 & 5680. & 107. &  0.27 & 0.05 &  4.06 & 0.26 \\
\enddata
\end{deluxetable}

\end{document}